\documentclass{moriond}

\def\Journal#1#2#3#4{{#1} {\bf #2}, #3 (#4)}

\def\PRL{\em Phys. Rev. Lett.}
\def\PRD{{\em Phys. Rev.} D}

\def\be{\begin{equation}}
\def\ee{\end{equation}}
\def\bea{\begin{eqnarray}}
\def\eea{\end{eqnarray}}

\newcommand{\Photo}{\includegraphics[height=0mm,angle=0]{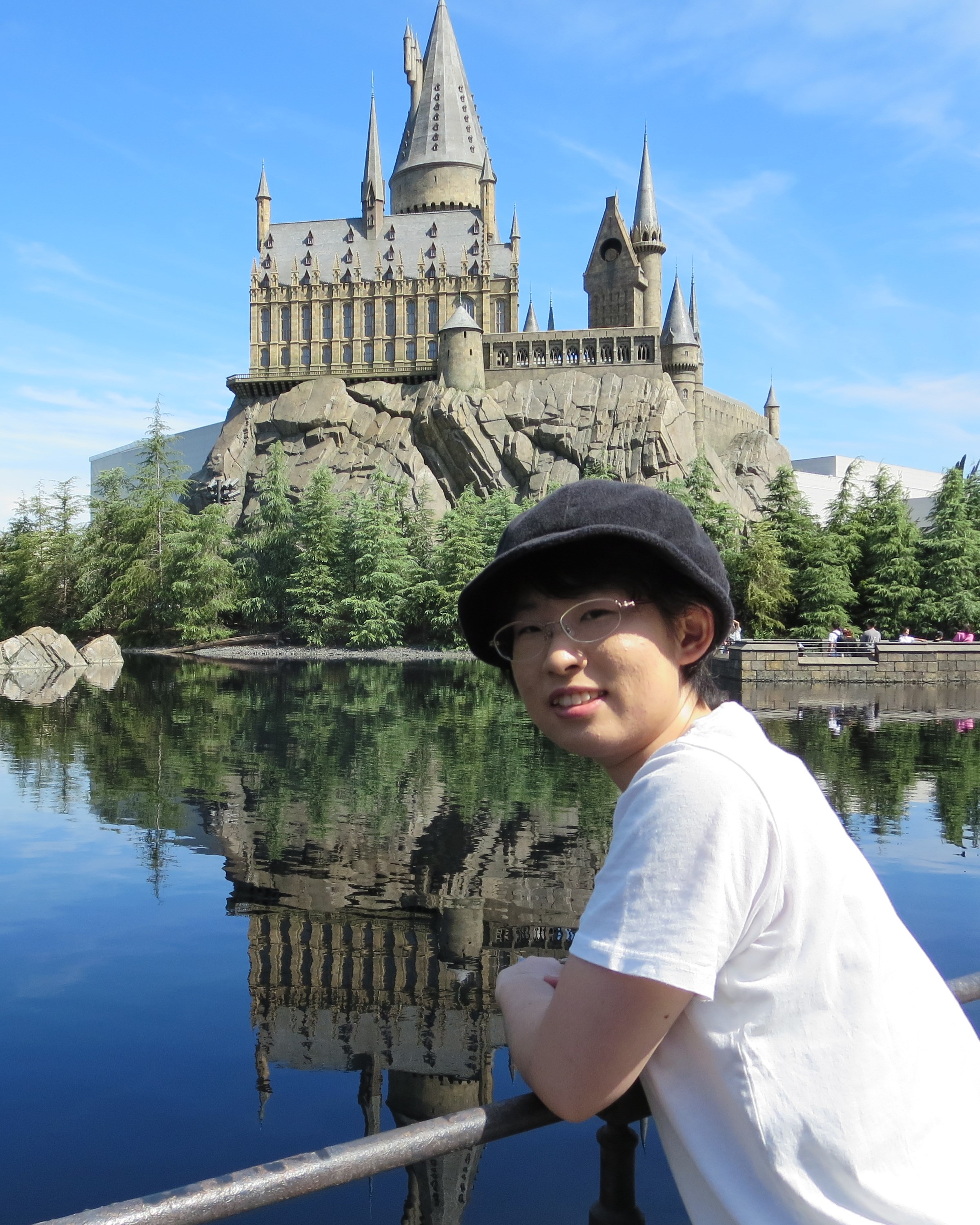}}

\begin{document}
\vspace*{1cm}
\title{Dark matter Axion search with riNg Cavity Experiment
\\DANCE: Current sensitivity}

\author{Yuka Oshima,$^1$ Hiroki Fujimoto,$^1$ Masaki Ando,$^1$ Tomohiro Fujita,$^2$\\
Yuta Michimura,$^1$ Koji Nagano,$^3$ Ippei Obata,$^4$ and Taihei Watanabe$^1$}

\vspace{10mm}

\address{$^1$Department of Physics, University of Tokyo, Bunkyo, Tokyo 113-0033, Japan\\
$^2$Institute for Cosmic Ray Research, University of Tokyo, Kashiwa 277-8582, Japan\\
$^3$Institute of Space and Astronautical Science, Japan Aerospace Exploration Agency, Sagamihara, Kanagawa 252-5210, Japan\\
$^4$Max-Planck-Institut f\"{u}r Astrophysik, Karl-Schwarzschild-Stra\ss e 1, 85741 Garching, Germany}

\maketitle\abstracts{
Dark matter Axion search with riNg Cavity Experiment (DANCE) was proposed.
To search for axion-like particles,
we aim to detect the rotation and oscillation of optical linear polarization
caused by axion-photon coupling with a bow-tie cavity.
DANCE can improve the sensitivity 
to axion-photon coupling constant $g_{a \gamma}$ 
for axion mass $m_a < 10^{-10} \, \rm{eV}$ 
by several orders of magnitude
compared to the best upper limits at present.
A prototype experiment DANCE Act-1 is in progress
to demonstrate the feasibility of the method and
to investigate technical noises.
We assembled the optics, evaluated the performance of the cavity,
and estimated the current sensitivity.
If we observe for a year, 
we can reach $g_{a \gamma} \simeq 9 \times 10^{-7} \, \rm{GeV^{-1}}$ 
at $m_a \simeq 10^{-13} \, \rm{eV}$.
The current sensitivity was believed to be limited
by laser intensity noise at low frequencies
and by mechanical vibration at high frequencies.
}

\section{Introduction}

Axions are pseudo-scalar fields 
originally proposed to solve the strong CP problem in QCD physics.
Moreover, string theory and supergravity generically predict
a plenitude of axion-like particles.
Hereafter we collectively call them “axions”.
Axions are one of the well-motivated candidates for dark matter
since axions typically have a small mass $m_a \ll \rm{eV}$ and
behave like non-relativistic classical wave fields
in the present universe
\cite{Fischler1} \cite{Fischler2} \cite{Fischler3} \cite{Obata}.
High energy physics predicts that
axions may weakly interact with photons~\cite{theory} \cite{theory2}.

A small coupling between axions and photons provides a good chance 
to detect axions through direct search experiments
by using well-developed photonics technology.
Recently, several novel methods were proposed
to observe axion-photon coupling
using carefully designed optical cavities
~\cite{linear} \cite{DANCE} \cite{ADBC} \cite{KAGRA} \cite{aux}.
These laser interferometric searches can be done
without a strong magnetic field,
and have good sensitivity in the low mass region.
In this paper, we review our
Dark matter Axion search with riNg Cavity Experiment (DANCE) proposal,
and report the current sensitivity of the prototype experiment,
DANCE Act-1.

\section{Principle of DANCE}

The axion-photon interaction gives
a phase velocity difference between left- and
right-handed circularly polarized light~\cite{theory} \cite{theory2}.
The phase velocity difference 
$\delta c = |c_{\rm{L}} - c_{\rm{R}}| 
= \delta c_0 \sin(m_a t + \delta_{\tau} (t))$
with axion mass $m_a$ and a phase factor $\delta_{\tau} (t)$
for a wavelength of light
$\lambda = 2\pi /k$
is estimated to be
\be
\delta c_0 = \frac{g_{a \gamma} a_0 m_a}{k} 
= 1.8 \times 10^{-24} \left( \frac{\lambda}{1064 \, \rm{nm}} \right)
\left( \frac{g_{a \gamma}}{10^{-12} \, \rm{GeV}} \right).
\ee
Here, we assumed axion energy density equals local dark matter density, 
$\rho_a = m_a^2 a_0^2 /2 \simeq 0.3 \, \rm{GeV} / \rm{cm}^3$.

This phase difference between circular polarizations is
equivalent to a rotation of linearly polarized light~\cite{theory}.
Small signal sidebands are generated 
as a linearly polarized laser propagates
in the presence of axions~\cite{ADBC}.
Optical path length can be effectively increased using an optical cavity
and the amplitude of the sidebands is enhanced for detection.
The polarization flip upon
mirror reflection have to be taken into account
when designing the optical cavities.
A bow-tie ring cavity is proposed 
to prevent the linear polarization from inverting
since the laser beam is reflected twice at both ends~\cite{DANCE}.

The fundamental noise source of DANCE would be quantum shot noise.
The one-sided amplitude spectral density of the shot noise is given by
\be
\sqrt{S_{\rm{shot}} (\omega)} =
\sqrt{\frac{\hbar \lambda}{4\pi c P_{\rm{trans}}}
\left( \frac{1}{t_c^2} + \omega^2 \right)},
\label{eq:shotnoise}
\ee
where $\omega$ is the fourier angular frequency
and $P_{\rm{trans}}$ is the transmitted laser power.
The averaged storage time of the cavity $t_c$
is given by $t_c = L \mathcal{F}/(\pi c)$,
where $L$ is the cavity round-trip length
and $\mathcal{F}$ is the finesse.
Assuming $L = 10 \, \rm{m}$, $\mathcal{F} = 10^6$,
and $P_{\rm{trans}} = 100 \, \rm{W}$,
we can reach
$g_{a \gamma} \simeq 3 \times 10^{-16} \, \rm{GeV^{-1}}$
for $m_a < 10^{-16} \, \rm{eV}$
(see purple dotted line in Figure \ref{fig:axionbounds}).
Here, we set $\lambda = 1064 \, \rm{nm}$ 
and the integration time $T = 1 \, \rm{year}$.
Simultaneous resonance of both carrier and sidebands beams
is also important for good sensitivity at low frequencies.

\section{Experimental Setups of DANCE Act-1}

\begin{figure}
\begin{minipage}{0.5\linewidth}
\centerline{\includegraphics[clip,width=65mm]{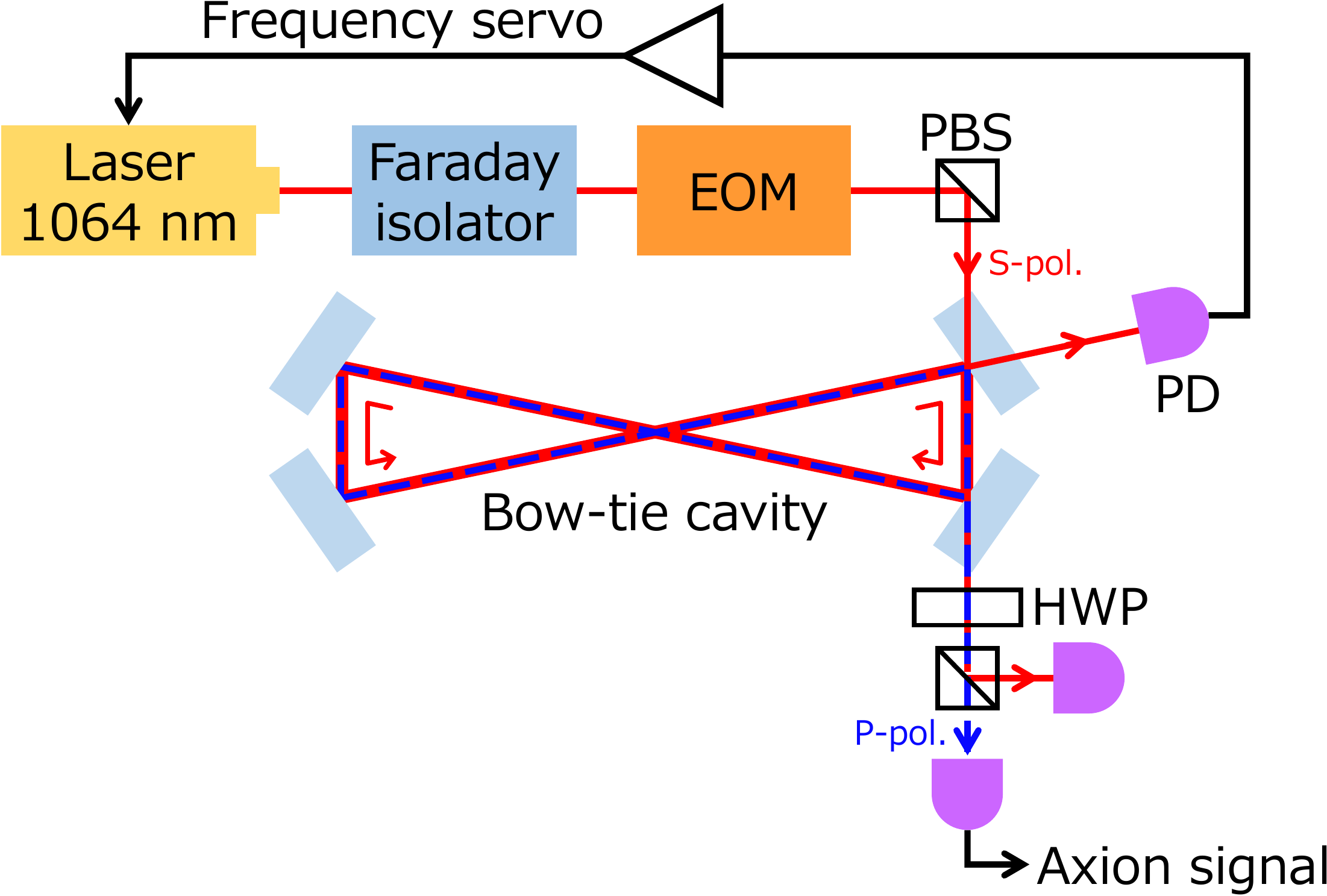}}
\end{minipage}
\hfill
\begin{minipage}{0.5\linewidth}
\centerline{\includegraphics[width=65mm]{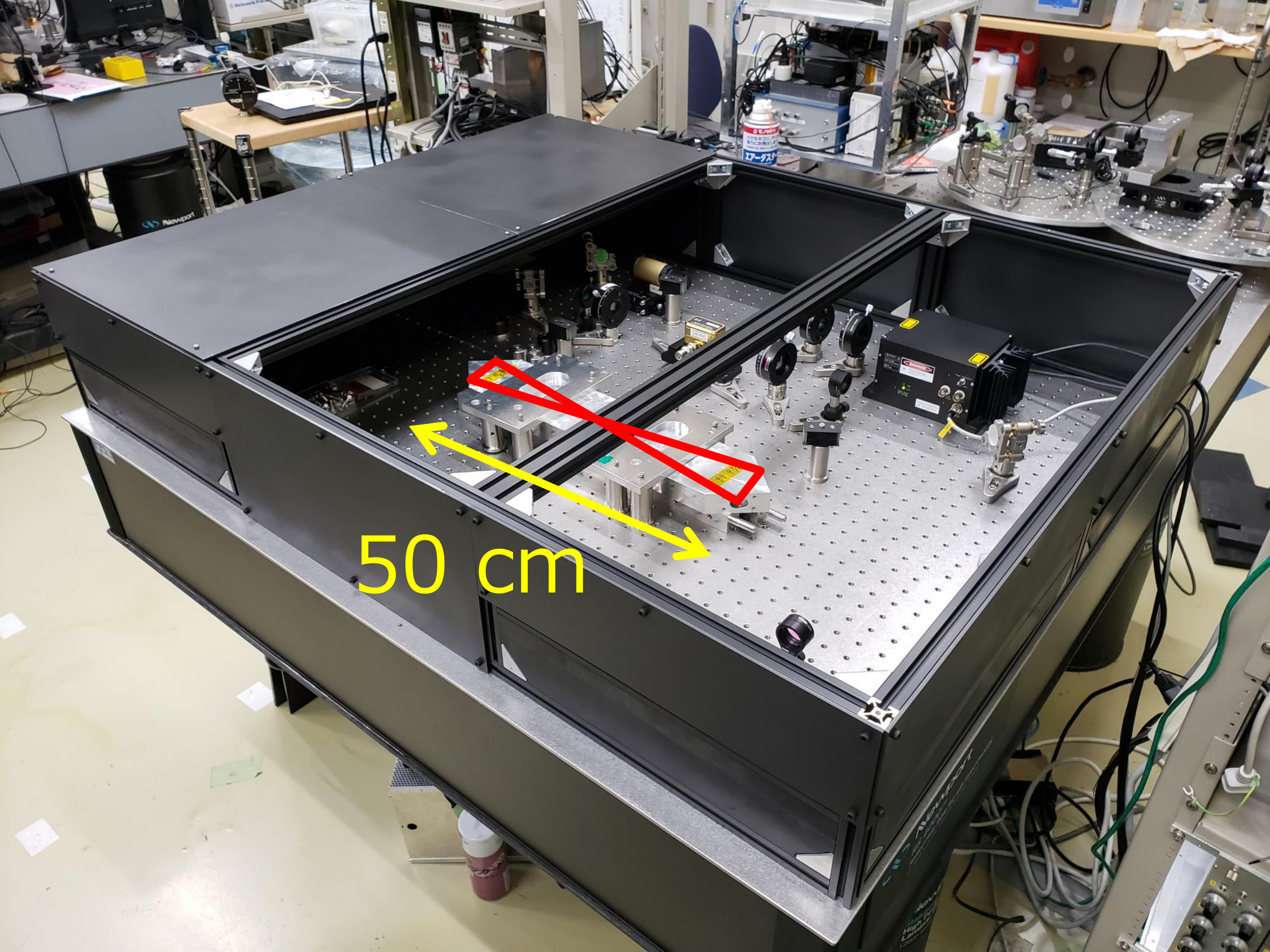}}
\end{minipage}
\caption[]{The schematic of DANCE Act-1 (left), 
and a picture of DANCE Act-1 (right).
S-polarized (P-polarized) beam is drawn as red (blue) lines.
EOM: electro-optic modulator.
PBS: polarizing beam splitter.
PD: photodetector.
HWP: half-wave plate.
}
\label{fig:setup}
\end{figure}

Figure \ref{fig:setup} (left) shows the schematic of DANCE Act-1.
The S-polarized beam (the carrier in this work)
was fed into the bow-tie cavity,
and the laser frequency was locked to the resonance
by the Pound-Drever-Hall technique.
Polarization of transmitted light was rotated with a half-wave plate (HWP)
to introduce some P-polarization (the sidebands in this work),
and then split
into S- and P-polarization with a polarizing beam splitter (PBS).
P-polarization can be measured with a photodetector (PD),
and in this signal we can search for axions.
The amount of S-polarization was also recorded with a PD
in order to calibrate the signal.

A photo of the experimental setup of DANCE Act-1
is shown in Figure \ref{fig:setup} (right).
The optical table was surrounded by aluminum plates
to stabilize frequency by reducing air turbulence
and shield the optical setup from external light.
The bow-tie cavity was constructed from four mirrors
rigidly fixed on a spacer made of aluminum.

\section{Results and Discussion of DANCE Act-1}

We evaluated the performance of the bow-tie cavity
by modulating the laser frequency 
and taking the cavity scan of transmitted light
(see Table \ref{tab:result}).
Measured values of round-trip length and finesse for S-polarization
were consistent with designed values.
Current injected laser power was lower than designed power
because we used a laser source with maximum power of $500 \, \rm{mW}$.
Transmitted laser power was lower than injected power
due to loss of light in the cavity.
Measured resonant frequencies for two polarizations were different
because S- and P-polarization obtained non-zero phase shift
from mirror coating layers
at mirror reflections with non-zero incident angle.
Measured finesse for P-polarization was smaller than the designed finesse
because the mirrors had lower reflectivity for P-polarization.
This was not an issue in this work
since the sensitivity in low mass region gets better with smaller finesse for P-polarization
when resonant frequency difference between polarizations is non-zero. 

\begin{table}
\caption[]{Summary of the performance evaluation of the bow-tie cavity.}
\label{tab:result}
\vspace{0.4cm}
\begin{center}
\begin{tabular}{|l|c|c|}
\hline
&
Designed values&
Measured values 
\\ \hline
Round-trip length $L$&
$99.4 \, \rm{cm}$&
$97.1(4.5) \, \rm{cm}$ 
\\ \hline
Injected laser power&
$1 \, \rm{W}$&
$274(14) \, \rm{mW}$ 
\\ \hline
Transmitted laser power $P_{\rm{trans}}$&
$1 \, \rm{W}$&
$158(8) \, \rm{mW}$ 
\\ \hline
Finesse for S-polarization (carrier) $\mathcal{F_{\rm{car}}}$&
$3 \times 10^3$&
$2.80(34) \times 10^3$ 
\\ \hline
Finesse for P-polarization (sidebands) $\mathcal{F_{\rm{side}}}$&
$3 \times 10^3$&
$193(10)$ 
\\ \hline
Resonant frequency difference between polarizations $\delta_{\rm{res}}$&
$0 \, \rm{Hz}$&
$3.92(16) \, \rm{MHz}$ 
\\ \hline
\end{tabular}
\end{center}
\end{table}

\begin{figure}
\begin{center}
\includegraphics[clip,width=0.8\linewidth]{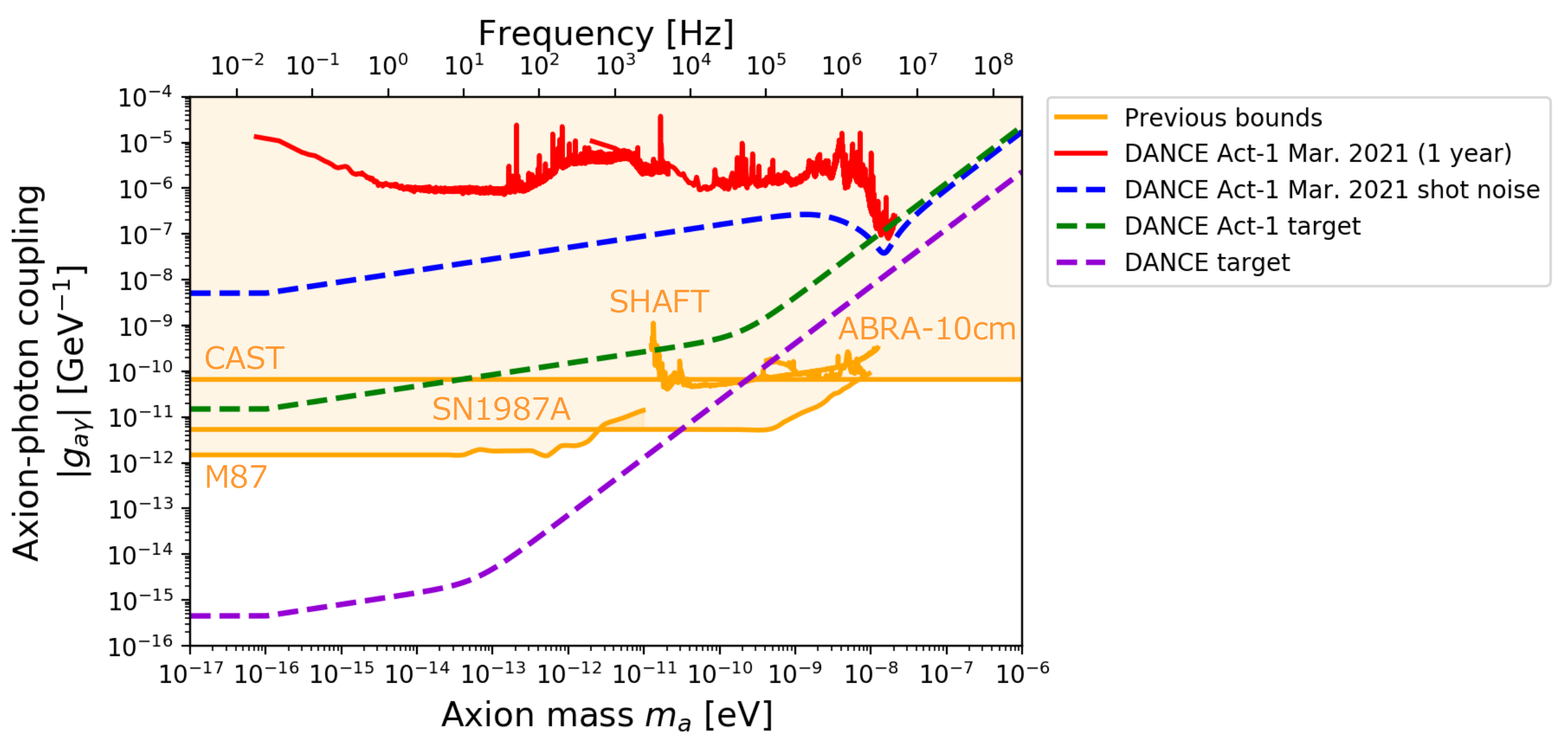}
\end{center}
\caption[]{The sensitivity curves 
for the axion-photon coupling constant $g_{a \gamma}$.
The orange solid lines with shaded region are current bounds obtained from
CAST~\cite{CAST} , SHAFT~\cite{SHAFT}, and ABRACADABRA-10cm~\cite{ABRA} experiments,
and the astrophysical constraints from
the gamma-ray observations of SN1987A~\cite{SN1987A} 
and the X-ray observations of M87 galaxy~\cite{M87}.
The red solid line shows the current estimated sensitivity of DANCE Act-1
if we observe for a year.
The dotted lines represent the expected shot noise limited sensitivity
of DANCE Act-1 with current setup parameters 
(blue; the cavity round-trip length of $L = 1 \, \rm{m}$,
transmitted laser power of $P_{\rm{trans}} = 158 \, {\rm{mW}}$,
finesse for carrier (S-polarization in this work) of
$\mathcal{F_{\rm{car}}} = 2.80 \times 10^3$,
finesse for sidebands (P-polarization in this work) of
$\mathcal{F_{\rm{side}}} = 193$,
and resonant frequency difference between polarizations of
$\delta_{\rm{res}} = 3.92 \, \rm{MHz}$),
DANCE Act-1 target (green; $L = 1 \, \rm{m}$,
$P_{\rm{trans}} = 1 \, {\rm{W}}$,
$\mathcal{F_{\rm{car}}} = \mathcal{F_{\rm{side}}} = 3 \times 10^3$,
and $\delta_{\rm{res}} = 0 \, \rm{Hz}$),
and DANCE target (purple; $L = 10 \, \rm{m}$,
$P_{\rm{trans}} = 100 \, {\rm{W}}$,
$\mathcal{F_{\rm{car}}} = \mathcal{F_{\rm{side}}} = 10^6$,
and $\delta_{\rm{res}} = 0 \, \rm{Hz}$)
with one-year integration time.
}
\label{fig:axionbounds}
\end{figure}

The amount of P-polarization $P_{\rm{P}} (t)$ was measured 
for 50 minutes with a PD and calibrated 
to the rotation angle of linear polarization $\phi (t)$ by 
\be
\phi (t) = \sqrt{\frac{P_{\rm{P}} (t)}{P_{\rm{tot}}}} - 2\theta,
\label{eq:rotangle}
\ee
where $P_{\rm{tot}}$ is the averaged total amount of transmitted light
and $\theta$ is the fixed angle of a HWP.
Then, the spectrum of the rotation angle of linear polarization
and the current estimated sensitivity were calculated.
If we observe for a year,
we can reach $g_{a \gamma} \simeq 9 \times 10^{-7} \, \rm{GeV^{-1}}$ 
at $m_a \simeq 10^{-13} \, \rm{eV}$
(see red solid line in Figure \ref{fig:axionbounds}).
Rotation angle of linear polarization
in $0.1 \, \rm{Hz}$ - $1 \, \rm{Hz}$
correlated significantly
with injected laser power,
therefore the current sensitivity was believed to be limited
by laser intensity noise.
Whereas, 
rotation angle of linear polarization
in $30 \, \rm{Hz}$ - $5 \, \rm{kHz}$
correlated significantly
with error signal for frequency servo,
therefore the current sensitivity was believed to be limited
by mechanical vibration.

\section{Conclusion}

A new table-top experiment, DANCE, was proposed
to search for axion dark matter with an optical cavity.
We aim to detect the rotation and oscillation of a linear polarization
caused by axion-photon coupling with a bow-tie cavity.
DANCE can improve the sensitivity
beyond the current bounds
of axion-photon coupling constant $g_{a \gamma}$
for axion mass $m_a < 10^{-10} \, \rm{eV}$
by several orders of magnitude.
A prototype experiment DANCE Act-1 is underway
to demonstrate the feasibility of the method.
We finished the assembly of the optics,
the performance evaluation of the cavity,
and the estimation of the current sensitivity.
If we observe for a year, 
we can reach $g_{a \gamma} \simeq 9 \times 10^{-7} \, \rm{GeV^{-1}}$ 
at $m_a \simeq 10^{-13} \, \rm{eV}$.
The current sensitivity was believed to be limited
by laser intensity noise at low frequencies
and by mechanical vibration at high frequencies.

We plan to observe for a week and analyze the data in May 2021.
Furthermore, we plan to build a new setup to improve the sensitivity
by injecting higher input laser power
and by canceling out resonant frequency difference
between polarizations with an auxiliary cavity~\cite{aux}.

\section*{Acknowledgments}

We would like to thank Shigemi Otsuka and Togo Shimozawa
for manufacturing the mechanical parts,
and Ching Pin Ooi for editing this paper.
This work was supported by JSPS KAKENHI Grant Nos. 18H01224,
20H05850, 20H05854 and 20H05859, and JST PRESTO Grant No. JPMJPR200B.


\section*{References}

\begin{thebibliography}{99}

\bibitem{Fischler1} J. Preskill, M. B. Wise and F. Wilczek,
\Journal{Phys. Lett. B}{120}{127}{1983}.

\bibitem{Fischler2} L. F. Abbott and P. Sikivie,
\Journal{Phys. Lett. B}{120}{133}{1983}.

\bibitem{Fischler3} M. Dine and W. Fischler,
\Journal{Phys. Lett. B}{120}{137}{1983}.

\bibitem{Obata} P. Arias {\it et al.},
\Journal{JCAP}{06}{013}{2012}.

\bibitem{theory} S. M. Carroll, G. B. Field and R. Jackiw,
\Journal{\PRD}{41}{1231}{1990}.

\bibitem{theory2} S. M. Carroll,
\Journal{\PRL}{81}{3067}{1998}.

\bibitem{linear} W. DeRocco and A. Hook,
\Journal{\PRD}{98}{035021}{2018}.

\bibitem{DANCE} I. Obata, T. Fujita and Y. Michimura, 
\Journal{\PRL}{121}{161301}{2018}.

\bibitem{ADBC} H. Liu, B. D. Elwood, M. Evans and J. Thaler,
\Journal{\PRD}{100}{023548}{2019}.

\bibitem{KAGRA} K. Nagano, T, Fujita, Y. Michimura and I. Obata,
\Journal{\PRL}{123}{111301}{2019}.

\bibitem{aux} D. Martynov and H. Miao, \Journal{\PRD}{101}{095034}{2020}.

\bibitem{CAST} CAST Collaboration, \Journal{Nature Physics}{13}{584}{2017}.

\bibitem{SHAFT} A. V. Gramolin {\it et al.},
\Journal{Nature Physics}{17}{79}{2021}.

\bibitem{ABRA} C. P. Salemi {\it et al.}, arXiv:2102.06722.

\bibitem{SN1987A} A. Payez {\it et al.},
\Journal{J. Cosmol. Astropart. Phys.}{02}{006}{2015}.

\bibitem{M87} M. C. David Marsh {\it et al.},
\Journal{J. Cosmol. Astropart. Phys.}{12}{036}{2017}.

\end{thebibliography}
\end{document}